\documentclass[12pt, aaspp4, flushrt, mathptmx]{aastex}

\usepackage{epsf}
\usepackage{rotate}

\newcommand{\ha}{H$\alpha$~}
\newcommand{\comments}[1]{}

\begin{document}

\sffamily

\title{Transverse Motions of Chromospheric Type II Spicules Observed by the New Solar Telescope}

\author{Yurchyshyn, V., Kilcik,  A., Abramenko, V.}

\affil{\it Big Bear Solar Observatory, New Jersey Institute of Technology, Big Bear City, CA 92314, USA}

\begin{abstract}

Using high resolution off-band \ha\ data from the New Solar Telescope and Morlet wavelet analysis technique, we analyzed transverse motions of type II spicules observed near the North Pole of the Sun. Our new findings are that i) some of the observed type II spicules display kink or an inverse ``Y'' features, suggesting that their origin may be due to magnetic reconnection, and ii) type II spicules tend to display coherent transverse motions/oscillations. Also, the wavelet analysis detected significant presence of high frequency oscillations in type II spicules, ranging from 30 to 180~s with the the average period of 90~s. We conclude that at least some of type II spicules and their coherent transverse motions may be caused by reconnection between large scale fields rooted in the intergranular lanes and and small-scale emerging dipoles, a process that is know to generate high frequency kink mode MHD waves propagating along the magnetic field lines.

\end{abstract}

\comments{Key words: Sun: atmosphere -- Sun: chromosphere -- Sun: oscillations}

\section{Introduction}

\noindent Spicules of type II \citep{2007PASJ...59S.655D} are ubiquitous chromospheric plasma upflows that are though to transfer of mass into the corona and be an important part of coronal heating and solar wind acceleration process \citep[e.g.,][]{2011Sci...331...55D,bart_roots,scot_upflows}. Type II spicules have a disk counterpart, the so called rapid blue-shifted excursions associated with the network fields \citep{langangen_2008, counterparts}.

Type II spicules (hereafter spicules II) possess physical properties different from classical spicules: a shorter lifetime (10~s -- 100~s), smaller width (150 -- 700~km) as well as much higher line-of-sight (50 -- 150~km s$^{-1}$) and transverse (10~km s$^{-1}$) velocities. Spectroscopic studies indicate that the plasma in these events is heated throughout their lifetime and the related upflows exhibit jet like properties \citep{2009ApJ...705..272R}.

Spicules II are also known to exhibit transverse/swaying, volume filling motions \citep[e.g.,][]{Tomczyk_2007, Scott_2011Natur} with amplitudes of 10-30~km s$^{-1}$ and periods of 100-500~s (see review by \cite{2009SSRv..149..355Z}), which are interpreted as upward or downward propagating ``Alfvenic'' waves \citep{Okamoto_Bart}, or MHD kink mode waves \citep[e.g.,][]{2009ApJ...705L.217H, 0004-637X-749-1-30, Kuridze_2012}.

There is little consensus among researchers as to how spicules II originate and what is the source of their transverse oscillations.  While some works suggest that reconnection process \citep{2008ApJ...679L..57I,2007PASJ...59S.655D,archontis_jet_model} and the oscillatory reconnection \citep{0004-637X-749-1-30} in particular may account for their origin, others propose that strong Lorentz force \citep{2011ApJ...736....9M} or propagation of the p-modes \citep{2009ApJ...702L.168D} may be at the origin. Moreover, \cite{2011ApJ...730L...4J} argues that spicules II could be warps in 2D sheet like structures (as opposed to tube-like structures), while \cite{zhang12revision} questions the existence of spicules II as a distinct class altogether. The related difficulties in interpretation of solar data mainly arise from limited spatial resolution and the complexity of the chromosphere \citep[e.g.,][]{2011ApJ...730L...4J,0004-637X-749-2-136}.

Although a substantial amount of work has been done in the field, the majority of studies dealt with off-limb data when data interpretation may be influenced by effects of line of sight integration and event selectivity, since only tallest and the most prominent spicules can be reliably measured. In particular, \cite{Moortel_2012} notes that the superposition of coronal loops may impair identification of both the oscillating structure and the mode of the observed oscillation. 

Here we focus on oscillations of type II spicules observed against the solar disk near the North Pole. In Section 3 we present results of wavelet analysis of off-band H$\alpha$ data obtained with the New Solar Telescope \citep[NST,][]{goode_nst_2010, Cao_IRIM} installed at the Big Bear Solar Observatory (BBSO), as well as discuss evolution of two spicule II events that exhibited signatures of a kink and inverted ``Y'' configuration.

\section{Observations, Data Reduction and Analysis}

\noindent High spatial resolution, 10 sec cadence NST data were taken near the solar limb with the aid of the Narrowband Filter Imager (NFI) under good seeing condition on September 8, 2011 between 17:41 and 19:29 UT. The NFI utilized a Zeiss Lyot filter with a 0.25\AA\ bandpass shifted relative the \ha line center by 1.0\AA\ toward the blue side of the spectra. Observing type II spicules at the far blue wing of the \ha\ spectral line also enables spatial filtering so that only high speed upflows of order of 50~km s$^{-1}$ directed toward the observed will appear in the field of view (FOV). This drastically reduces the number of detected absorption features in the FOV thus enabling us to better resolve and track individual events. The optical setup included a 1.6~m primary mirror and an adaptive optics system with a 97 actuator deformable mirror. The data set was intermittent with numerous small data gaps, therefore we only used one longest (34 min) continuous data segment consisting of 192 images (pixel scale 0''.052) that were speckle reconstructed \citep{kisip_code}, aligned and de-stretched to remove the residual image distortion due to seeing and image jittering. 

\begin{figure}[!th]
\centerline{\epsfxsize=6truein \epsffile{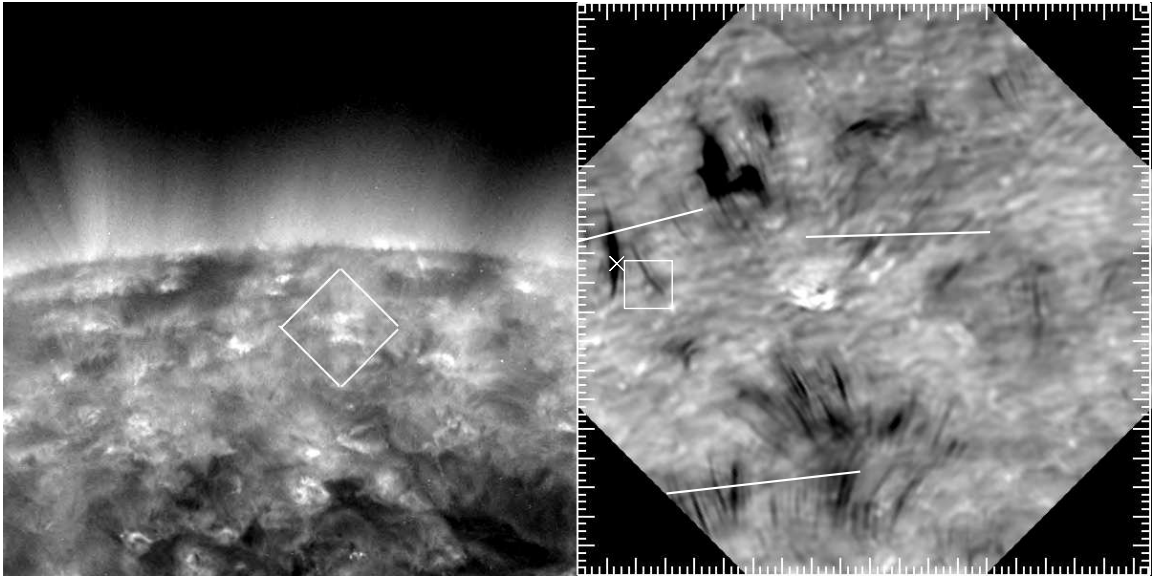}}
\caption{SDO/AIA 171\AA\ (left) and H$\alpha$-1.0\AA\ (right) images taken at 17:59~UT on September 8, 2011. The white square in the left panel shows position and size of the entire FOV of the NST \ha\ data (38 $\times$ 38~Mm or 52'' $\times$ 52''). In order to show details only the central part (20 $\times$ 20~Mm) of the entire \ha\ FOV is shown in the right panel. The box in the right panel encloses features analysed in this paper. The three line segments indicate the locations along which the x-t plots in Figure \ref{xtplot} were produced, and the white ``x'' indicates the position of the intensity profile in Figure \ref{profile}. The tickmarks in the right panel separate 1~Mm intervals. The North Pole is near the upper left corner of the right panel. (An animation (S1.mpeg) of the right panel is available in the online journal.)}
\label{data}
\end{figure}

In Figure \ref{data} we present the data set and its location relative to the solar limb. The image on the left is a cut out of the full disk ADO/AIA 171\AA\ image with the white diamond indicating the FOV of the \ha-1.0\AA\ image on the right. On the day of observations the solar B$_0$ angle was maximal (7.25~deg), so that the NST/\ha\ FOV of 52'', centered at heliocentric coordinates 31''902'', included a quiet Sun area near the North Pole (heliocentric coordinates 0''945''). According to the SDO/AIA image, the polar region and the FOV was not occupied by a coronal hole at this time.

In the center of the FOV there was a small distinct cluster of bright points Figure \ref{data}, right) that was the focal point of a large number of dynamic jet-like features, which showed various oscillating and swaying motions (S1.mpg).  We, however, will focus on activities to the left of the cluster marked by various symbols.  Since these jets are similar to the rapid blue-shifted excursions identified as type II spicules \citep{2008ApJ...679L.167L, counterparts}, we accept here that they are type II spicules. One large group of spicules II was located in the lower right corner of the right panel in Figure \ref{data}. Another group of active spicules II was located above and to the left of the black box in the upper left corner. Interestingly, the same locations in the SDO/AIA 171\AA\ image (left panel) show enhanced bright structures that often indicate closed magnetic loops connecting network fields. It therefore appears that a significant number of the spicules II could be associated with relatively short closed loops.

\begin{figure}[!p]
\centerline{\epsfxsize=4.7truein  \epsffile{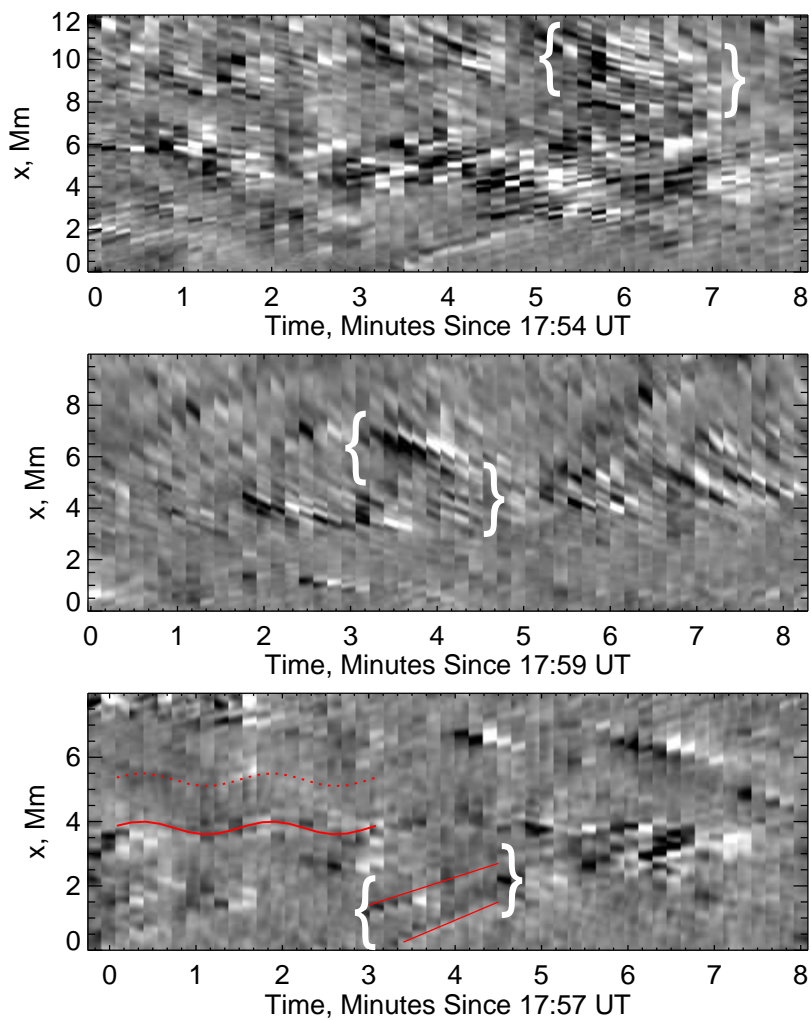}}
\caption{x-t plots made along the cuts in Figure \ref{data}. To produce these plots we used \ha difference images, which were generated by subtracting 2 consecutive images, and then cutting out strips of 0.3~Mm wide along the line segments and plotting them after each other. The top x-t plot was made along the lower cut in Figure \ref{data} (rigth), the middle plot is from the rightmost cut and the bottom x-t plot is from the leftmost cut. The parenthesis and the two line segments enclose several coherently displaced features. The solid sinusoidal line is the best fit of an harmonic oscillator to the wavelike displacements of the x-t features. These nearly two full oscillations had a period of 90~s, amplitude of about 200~km and the amplitude velocity of about 9~km s$^{-1}$. The dotted line is the same sinusoidal fit displaced in the vertical direction by about 1.5 Mm. The pattern of the x-t plot between those two lines suggests that there were multiple very faint type II spicules that appeared to coherently oscillate.}
\label{xtplot}
\end{figure}

\section{Results}

In Figure \ref{xtplot}, we show three x-t plots made across type II spicules (white lines in Figure \ref{data}). In order to enhance the contrast of the features, the x-t plots were produced using \ha difference images (see details in figure captions). It is interesting to note that the phase of displacements of many white and black streaks along the vertical $x$-axis (e.g., inclined black and white ridges between parenthesis) seems to be similar, implying that a group of neighbouring spicules may undergo coherent transverse displacement. The two top panels show numerous instances of such coherent transverse motions as evidenced by the linear tracks. The inclination of the ridges suggests that these chromospheric structures were displaced with speeds of order of 15~km s$^{-1}$, which corresponds to Alfv{\'e}n velocity in the chromosphere.

In the lower panel we show an example of coherent high frequency oscillations of type II spicules. The white solid line is a sinusoidal (harmonic oscillator) fit to the observed wave. The period of the wave was found to be around 90~s, and the amplitude of about 0.2~Mm, giving the amplitude velocity of about 6~km s$^{-1}$. The dotted line is the same sinusoidal fit displaced in the vertical direction by about 1.5~Mm.  The wavy pattern of the x-t plot between those two lines suggests that there were multiple very faint type II spicules that appeared to coherently oscillate.

\begin{figure}[!h]
\centerline{\epsfxsize=5truein  \epsffile{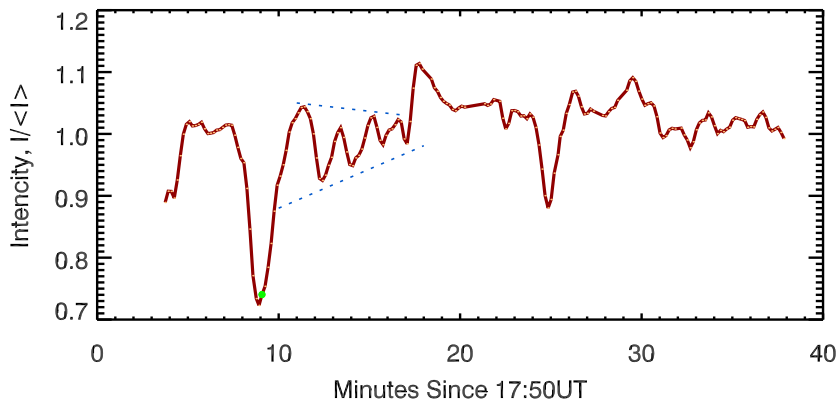}}
\caption{Normalized intensity variations at the ``x'' spot in the right panel of Figure \ref{data}. The green spot on the profile located at 17:59~UT indicates the time of the \ha images in Figures \ref{data} and \ref{wavelet}, while the two blue dotted lines highlight intensity oscillations observed after a dark and large absorption feature appeared in the field of view.}
\label{profile}
\end{figure}

As a dark absorption feature sways across the FOV, the intensity at a given pixel will fluctuate with the period of the swaying motions. The intensity profile in Figure \ref{profile} was derived at the location marked by the ``x'' in Figure \ref{data}. These fluctuations were associated with appearance of a dark feature at 17:58~UT, which showed 4 full periods (P=1.5~min) of swaying motions. We emphasize that the damped oscillations between the blue dotted lines represent intensity oscillations of a gradually fading absorption feature rather than damping of the swaying motions of the entire structure.

To understand the spatial distribution of these intensity oscillations we used the wavelet analysis technique, which has been previously applied to solar images. In particular, \cite{roupe_2007} studied the dynamics of quiet sun mottles. For the analysis we used Morlet wavelet with the red noise approximation \citep{torrence&compo}. As a tested variable we used 122500 intensity time profiles derived from the data cube by averaging pixel values over a 2$\times$2 pixel area. The wavelet analysis was applied to each intensity profile and the period with highest significance level was then defined as the dominant period. The significance was calculated by normalizing the period power with the power of 95\% confidence level of the global wavelet spectrum. A period was considered significant and used for analysis (with 95\% of significance level) when this ratio exceeds unity \citep[see][]{torrence&compo}.

\begin{figure}[!h]
\centerline{\epsfxsize=6.5truein  \epsffile{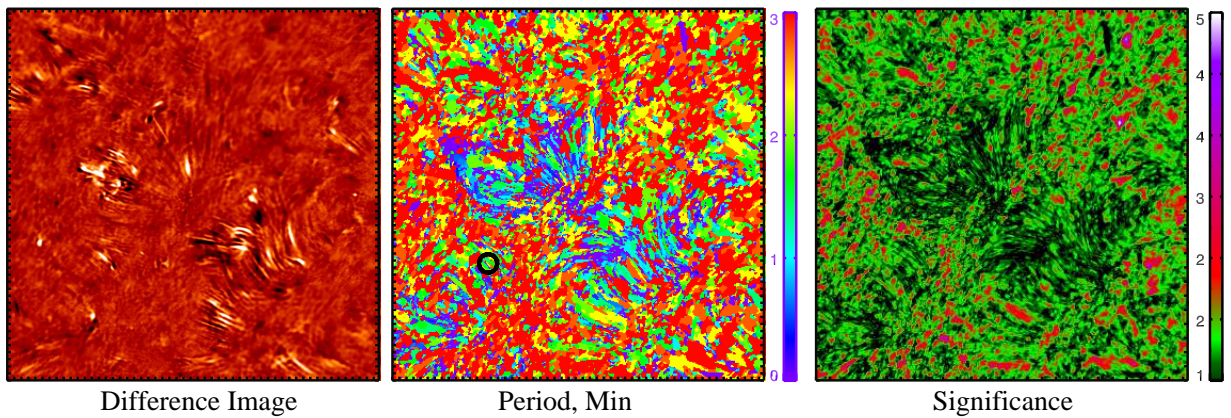}}
\caption{\ha difference image (left) and the corresponding maps of the dominant wavelet periods (middle) and their significance level (right). The circle plotted over the the significance map encloses the locality associated with the sinusoidal wave in Figure \ref{xtplot}. }
\label{wavelet}
\end{figure}

Results of the global wavelet analysis are shown in Figure \ref{wavelet}. The general pattern of the spatial distribution of  spicules II seen in the original data is also evident in maps of the dominant period (middle) and significance (right). Comparing the panels one may conclude that short period oscillations ($<2$~min, green-blue hue in the period map) mainly emanate from the area of spicules II, while longer period oscillations (yellow and red) come from quiet sun areas outside the domain of type II spicules. We also would like to point out that the wavelet analysis at the location marked by the circle in the middle panel of Figure \ref{wavelet} and the harmonic oscillator fit to the corresponding wave in the x-t plot (Figure \ref{xtplot}, lower panel) produce very close periods of about 90~s. Generally speaking,  the intensity profiles, analysed by the wavelet technique, are the result of multiple swaying structures that appear and fade at the same location. The dominant wavelet periods, discussed above, describe the most powerful single oscillation event that occurred during the studied time interval, rather than present a cumulative power of a certain period spread over the time interval.

What is the driver of the swaying motions that spicules II display? Their association with high frequency waves ($>$8~mHz) seem to indicate that they are unlikely to be driven by the photospheric $p$-mode oscillations. Based on the MHD simulations, \cite{2008ApJ...679L..57I} showed that reconnection of an emerging flux with pre-existing fields may produce high-frequency waves that propagate into the corona. The simulated transverse velocities associated with the reconnection region oscillate with a period of about 90~s, which is the average period detected in our study. 

\begin{figure}[!h]
\centerline{\epsfxsize=5truein\epsffile{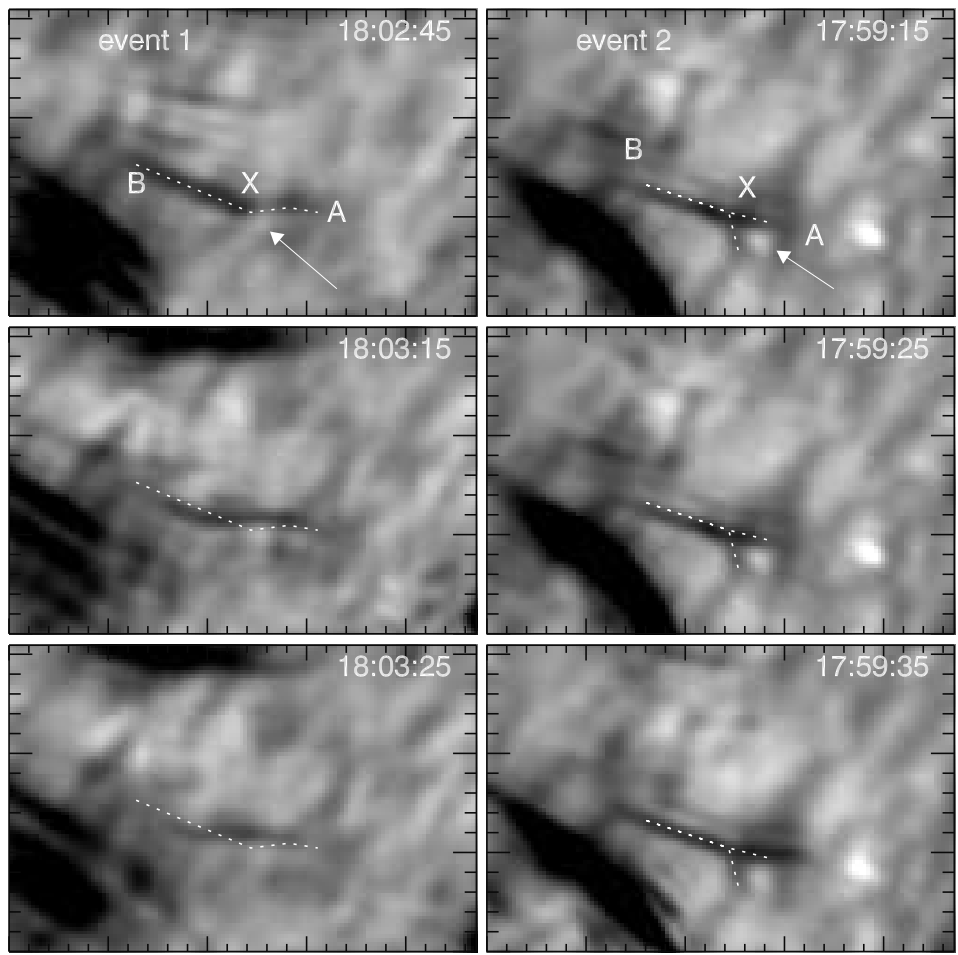}}
\caption{Evolution of two spicule II events that exhibit signatures of reconnection (e.g., kink and inverted ``Y'' shapes). These jets were rooted at point A, the point X may signify the field lines bending above the reconnection region, while the arc between A and X may be overlaying the emerging dipole. Point B marks the dynamic top part of the jet event. The dotted line in each panel represents the shape of the feature as captured in the top panel. The arrows point indicate small and weak brightenings associated with these jets. The large tick marks separate 1~Mm intervals.}
\label{case}
\end{figure}

In Figure \ref{case} we show two jets that exhibit signatures of magnetic reconnection. The shape of these jets is identical to that produced in simulations of reconnection of emerging flux and pre-existing large-scale fields (see e.g., Figure 3 in \cite{shibata2007} and Figure 1 in \cite{2008ApJ...679L..57I}). In both cases jets appeared rooted at point A. Point X marks the location where field lines bend above the reconnection point, and the arc between A and X is spanning the emerging dipole. As event 1 evolved, the X point as well as its tip (point B) moved away from its original position (compare the dotted line with the evolving shape of the features). At about 18:02~UT a small low contrast brightening (arrow in the top panel) developed just below X. Shortly after, point B retracted back toward X and the jet faded from the FOV.  Event 2 was an inverted ``Y'' shaped jet, which is a signature of reconnection as well \citep{shibata2007}. It was similarly associated with a weak brightening (arrow in the top panel). This event did not appear as dynamics as event 1 and only leg separation of the inverted ``Y'' structure slightly increased with time.  Both events faded from the FOV some 20~s after the time of the last corresponding frames in Figure \ref{case}.

These kink and inverted ``Y'' shaped jets clearly show that at least some of the type II spicules may be of reconnection origin. The so called interchange reconnection between an emerging closed loops and ``open'' field line may lead to a rapid displacement of a magnetic footpoint and thus generating high frequency kink mode MHD waves propagating along the magnetic field lines. 

\section{Discussion and Conclusions}

Using high resolution images of spicules II and Morlet wavelet analysis technique, we analysed transverse motions of type II spicules observed by the NST at the North Pole regions. The data suggests that the spicule II activity quite often exhibit coherent transverse motions. The width of the band of coherently moving spicules is about 1.5-2.0~Mm. We report periods of the transverse motions as detected by the wavelet technique in the range of 30-180~s with the most probable period of about 90~s. MHD simulations by \cite{2008ApJ...679L..57I} show that such high frequency waves with periods of 90~s are likely to be driven by small-scale reconnection processes. They also could be the result of photospheric oscillations leaking in to the chromosphere, since the detected periods are below the cut-off periods of chromospheric oscillations and kink-waves \citep{Kuridze_2012}. Similar short period transverse motions were recently detected by \cite{Okamoto_Bart} using of limb observations in the Ca II H line \citep[see also ][]{2009SSRv..149..355Z, 2007Sci...318.1574D, Scott_2011Natur}. Present wavelet results emphasise significant presence of short period oscillations associated with type II spicules.

We also found that some spicules of type II display kink and inverted ``Y'' shape. These features were detected at the very root of spicules II, where they probably originate. We thus argue that at least some spicules II events and their transverse motions may be due to the interchange type of reconnection between the ``open'' (or large scale closed) field lines and the emerging small-scale dipoles accompanied by generation of high frequency kink mode MHD waves propagating along the magnetic field lines. At a first glance, the interchange reconnection approach may not be agreeing with the coherent oscillations of spicules. One possibility is that a reconnection event removes flux from a bundle of flux tubes thus inducing rapid shuffling of footpoints and displacement of flux tubes within a cluster as this suddenly disturbed magnetic system undergoes equilibrium reconfiguration. The disturbance may generate both small scale (component) reconnection and high frequency MHD waves as proposed by \cite{0004-637X-736-1-3}, which may account for the appearance of smaller and thinner spicules II as well as their group oscillations. 

The majority of existing off-limb studies were focused on tallest and most prominent features, possibly associated with ``open'' and/or closed large loops that tend to oscillate with lower frequencies due to their larger extend and lower intensity of the magnetic field \citep[see e.g.,][]{2001A&A...372L..53N}. However, it appears that the majority of the spicules II in our data set appears to be associated with closed loops that probably span one or two super-granular cells (30-60~Mm). Their fields on, average, may be stronger and the observed transverse motions may be dominated by shorter period oscillations. Moreover, the number density of spicules II in our data set is much lower that that in the off-limb data, so we had an opportunity to resolve and more accurately detect individual events. For instance, we reliably measure transverse motions of these well resolved \ha\ structures at heights below 3-5~Mm. While we detect in many cases at least one period of oscillations, the off-limb data at these heights tend to produce linear tracks \citep{2007Sci...318.1574D} that may be less reliable in determining periods. Combination of these facts may be the reason  that the short period oscillations are prominently present in our study \citep[also see relevant conclusions by][]{Moortel_2012}.

This work was conducted as part of the NASA Living with a Star ``Jets'' Focused Science Team. Authors thank T.~J. Okamoto, B. de Pontieu, T.~J. Wang and H. Tian for valuable discussions. Authors acknowledge the usage of the Heliophysics Events Knowledgebase and data from NASA SDO mission. We thank BBSO observing and engineering staff for support and observations. This research was supported by NASA grants GI NNX08AJ20G, LWS NNX08AQ89G,  NNX11AO73G, as well as NSF AGS-1146896 grant.


\begin{thebibliography}{33}
\expandafter\ifx\csname natexlab\endcsname\relax\def\natexlab#1{#1}\fi

\bibitem[{{Abramenko} {et~al.}(2011){Abramenko}, {Carbone}, {Yurchyshyn},
  {Goode}, {Stein}, {Lepreti}, {Capparelli}, \&
  {Vecchio}}]{2011ApJ...743..133A}
{Abramenko}, V.~I., {Carbone}, V., {Yurchyshyn}, V., {Goode}, P.~R., {Stein},
  R.~F., {Lepreti}, F., {Capparelli}, V., \& {Vecchio}, A. 2011, \apj, 743, 133

\bibitem[{{Archontis} {et~al.}(2010){Archontis}, {Tsinganos}, \&
  {Gontikakis}}]{archontis_jet_model}
{Archontis}, V., {Tsinganos}, K., \& {Gontikakis}, C. 2010, Astron. Astrophys.,
  512, L2

\bibitem[{{Cao} {et~al.}(2010){Cao}, {Gorceix}, {Coulter}, {Ahn}, {Rimmele}, \&
  {Goode}}]{Cao_IRIM}
{Cao}, W., {Gorceix}, N., {Coulter}, R., {Ahn}, K., {Rimmele}, T.~R., \&
  {Goode}, P.~R. 2010, Astronomische Nachrichten, 331, 636

\bibitem[{{De Pontieu} {et~al.}(2009){De Pontieu}, {McIntosh}, {Hansteen}, \&
  {Schrijver}}]{bart_roots}
{De Pontieu}, B., {McIntosh}, S.~W., {Hansteen}, V.~H., \& {Schrijver}, C.~J.
  2009, \apjl, 701, L1

\bibitem[{{de Pontieu} {et~al.}(2007){de Pontieu}, {McIntosh}, {Hansteen},
  {Carlsson}, {Schrijver}, {Tarbell}, {Title}, {Shine}, {Suematsu}, {Tsuneta},
  {Katsukawa}, {Ichimoto}, {Shimizu}, \& {Nagata}}]{2007PASJ...59S.655D}
{de Pontieu}, B., {et~al.} 2007, \pasj, 59, 655

\bibitem[{{De Pontieu} {et~al.}(2007){De Pontieu}, {McIntosh}, {Carlsson},
  {Hansteen}, {Tarbell}, {Schrijver}, {Title}, {Shine}, {Tsuneta}, {Katsukawa},
  {Ichimoto}, {Suematsu}, {Shimizu}, \& {Nagata}}]{2007Sci...318.1574D}
{De Pontieu}, B., {et~al.} 2007, Science, 318, 1574

\bibitem[{{De Pontieu} {et~al.}(2011){De Pontieu}, {McIntosh}, {Carlsson},
  {Hansteen}, {Tarbell}, {Boerner}, {Martinez-Sykora}, {Schrijver}, \&
  {Title}}]{2011Sci...331...55D}
---. 2011, Science, 331, 55

\bibitem[{{de Wijn} {et~al.}(2009){de Wijn}, {McIntosh}, \& {De
  Pontieu}}]{2009ApJ...702L.168D}
{de Wijn}, A.~G., {McIntosh}, S.~W., \& {De Pontieu}, B. 2009, \apjl, 702, L168

\bibitem[{{Goode} {et~al.}(2010){Goode}, {Coulter}, {Gorceix}, {Yurchyshyn}, \&
  {Cao}}]{goode_nst_2010}
{Goode}, P.~R., {Coulter}, R., {Gorceix}, N., {Yurchyshyn}, V., \& {Cao}, W.
  2010, Astronomische Nachrichten, 88, 789

\bibitem[{{He} {et~al.}(2009){He}, {Marsch}, {Tu}, \&
  {Tian}}]{2009ApJ...705L.217H}
{He}, J., {Marsch}, E., {Tu}, C., \& {Tian}, H. 2009, \apjl, 705, L217

\bibitem[{{Isobe} {et~al.}(2008){Isobe}, {Proctor}, \&
  {Weiss}}]{2008ApJ...679L..57I}
{Isobe}, H., {Proctor}, M.~R.~E., \& {Weiss}, N.~O. 2008, \apjl, 679, L57

\bibitem[{{Judge} {et~al.}(2011){Judge}, {Tritschler}, \& {Chye
  Low}}]{2011ApJ...730L...4J}
{Judge}, P.~G., {Tritschler}, A., \& {Chye Low}, B. 2011, \apjl, 730, L4

\bibitem[{{Kuridze} {et~al.}(2012){Kuridze}, {Morton}, {Erd{\'e}lyi},
  {Dorrian}, {Mathioudakis}, {Jess}, \& {Keenan}}]{Kuridze_2012}
{Kuridze}, D., {Morton}, R.~J., {Erd{\'e}lyi}, R., {Dorrian}, G.~D.,
  {Mathioudakis}, M., {Jess}, D.~B., \& {Keenan}, F.~P. 2012, \apj, 750, 51

\bibitem[{{Langangen} {et~al.}(2008{\natexlab{a}}){Langangen}, {De Pontieu},
  {Carlsson}, {Hansteen}, {Cauzzi}, \& {Reardon}}]{langangen_2008}
{Langangen}, {\O}., {De Pontieu}, B., {Carlsson}, M., {Hansteen}, V.~H.,
  {Cauzzi}, G., \& {Reardon}, K. 2008{\natexlab{a}}, \apjl, 679, L167

\bibitem[{{Langangen} {et~al.}(2008{\natexlab{b}}){Langangen}, {De Pontieu},
  {Carlsson}, {Hansteen}, {Cauzzi}, \& {Reardon}}]{2008ApJ...679L.167L}
---. 2008{\natexlab{b}}, \apjl, 679, L167

\bibitem[{Leenaarts {et~al.}(2012)Leenaarts, Carlsson, \& van~der
  Voort}]{0004-637X-749-2-136}
Leenaarts, J., Carlsson, M., \& van~der Voort, L.~R. 2012, The Astrophysical
  Journal, 749, 136

\bibitem[{{Mart{\'{\i}}nez-Sykora} {et~al.}(2011){Mart{\'{\i}}nez-Sykora},
  {Hansteen}, \& {Moreno-Insertis}}]{2011ApJ...736....9M}
{Mart{\'{\i}}nez-Sykora}, J., {Hansteen}, V., \& {Moreno-Insertis}, F. 2011,
  \apj, 736, 9

\bibitem[{{McIntosh} \& {De Pontieu}(2009)}]{scot_upflows}
{McIntosh}, S.~W., \& {De Pontieu}, B. 2009, \apj, 707, 524

\bibitem[{{McIntosh} {et~al.}(2011){McIntosh}, {de Pontieu}, {Carlsson},
  {Hansteen}, {Boerner}, \& {Goossens}}]{Scott_2011Natur}
{McIntosh}, S.~W., {de Pontieu}, B., {Carlsson}, M., {Hansteen}, V., {Boerner},
  P., \& {Goossens}, M. 2011, \nat, 475, 477

\bibitem[{{McLaughlin} {et~al.}(2012){McLaughlin}, {Verth}, {Fedun}, \&
  {Erd{\'e}lyi}}]{0004-637X-749-1-30}
{McLaughlin}, J.~A., {Verth}, G., {Fedun}, V., \& {Erd{\'e}lyi}, R. 2012, The
  Astrophysical Journal, 749, 30

\bibitem[{Moortel \& Pascoe(2012)}]{Moortel_2012}
Moortel, I.~D., \& Pascoe, D.~J. 2012, The Astrophysical Journal, 746, 31

\bibitem[{{Nakariakov} \& {Ofman}(2001)}]{2001A&A...372L..53N}
{Nakariakov}, V.~M., \& {Ofman}, L. 2001, \aap, 372, L53

\bibitem[{{Okamoto} \& {De Pontieu}(2011)}]{Okamoto_Bart}
{Okamoto}, T.~J., \& {De Pontieu}, B. 2011, \apjl, 736, L24

\bibitem[{{Rouppe van der Voort} {et~al.}(2009{\natexlab{a}}){Rouppe van der
  Voort}, {Leenaarts}, {de Pontieu}, {Carlsson}, \& {Vissers}}]{counterparts}
{Rouppe van der Voort}, L., {Leenaarts}, J., {de Pontieu}, B., {Carlsson}, M.,
  \& {Vissers}, G. 2009{\natexlab{a}}, \apj, 705, 272

\bibitem[{{Rouppe van der Voort} {et~al.}(2009{\natexlab{b}}){Rouppe van der
  Voort}, {Leenaarts}, {de Pontieu}, {Carlsson}, \&
  {Vissers}}]{2009ApJ...705..272R}
---. 2009{\natexlab{b}}, \apj, 705, 272

\bibitem[{{Rouppe van der Voort} {et~al.}(2007){Rouppe van der Voort}, {De
  Pontieu}, {Hansteen}, {Carlsson}, \& {van Noort}}]{roupe_2007}
{Rouppe van der Voort}, L.~H.~M., {De Pontieu}, B., {Hansteen}, V.~H.,
  {Carlsson}, M., \& {van Noort}, M. 2007, \apjl, 660, L169

\bibitem[{{Shibata} {et~al.}(2007){Shibata}, {Nakamura}, {Matsumoto}, {Otsuji},
  {Okamoto}, {Nishizuka}, {Kawate}, {Watanabe}, {Nagata}, {UeNo}, {Kitai},
  {Nozawa}, {Tsuneta}, {Suematsu}, {Ichimoto}, {Shimizu}, {Katsukawa},
  {Tarbell}, {Berger}, {Lites}, {Shine}, \& {Title}}]{shibata2007}
{Shibata}, K., {et~al.} 2007, Science, 318, 1591

\bibitem[{{Tomczyk} {et~al.}(2007){Tomczyk}, {McIntosh}, {Keil}, {Judge},
  {Schad}, {Seeley}, \& {Edmondson}}]{Tomczyk_2007}
{Tomczyk}, S., {McIntosh}, S.~W., {Keil}, S.~L., {Judge}, P.~G., {Schad}, T.,
  {Seeley}, D.~H., \& {Edmondson}, J. 2007, Science, 317, 1192

\bibitem[{{Torrence} \& {Compo}(1998)}]{torrence&compo}
{Torrence}, C., \& {Compo}, G. 1998, 61

\bibitem[{van Ballegooijen {et~al.}(2011)van Ballegooijen, Asgari-Targhi,
  Cranmer, \& DeLuca}]{0004-637X-736-1-3}
van Ballegooijen, A.~A., Asgari-Targhi, M., Cranmer, S.~R., \& DeLuca, E.~E.
  2011, The Astrophysical Journal, 736, 3

\bibitem[{{W{\"o}ger} \& {von der L{\"u}he}(2007)}]{kisip_code}
{W{\"o}ger}, F., \& {von der L{\"u}he}, O. 2007, Appl. Opt., 46, 8015

\bibitem[{{Zaqarashvili} \& {Erd{\'e}lyi}(2009)}]{2009SSRv..149..355Z}
{Zaqarashvili}, T.~V., \& {Erd{\'e}lyi}, R. 2009, \ssr, 149, 355

\bibitem[{{Zhang} {et~al.}(2012){Zhang}, {Shibata}, {Wang}, {Mao}, {Matsumoto},
  {Liu}, \& {Su}}]{zhang12revision}
{Zhang}, Y.~Z., {Shibata}, K., {Wang}, J.~X., {Mao}, X.~J., {Matsumoto}, T.,
  {Liu}, Y., \& {Su}, J.~T. 2012, ArXiv e-prints

\end{thebibliography}

\end{document}